\documentclass[11pt,a4paper]{article}
\usepackage[utf8]{inputenc}
\usepackage[T1]{fontenc}
\usepackage[english]{babel}
\usepackage{lmodern}              
\usepackage{amsmath, amssymb, amsfonts} 
\usepackage{algorithm}
\usepackage{algorithmic}
\usepackage{siunitx}              
\usepackage{physics}              
\usepackage{bm}                   
\usepackage{graphicx}             
\usepackage{caption}
\usepackage{subcaption}           
\usepackage{booktabs}             
\usepackage{geometry}             
\usepackage{microtype}            
\usepackage{authblk}              
\usepackage{hyperref}             
\usepackage{xcolor}               
\usepackage{setspace}             
\usepackage{tikz}

\usepackage[backend=biber,style=ieee,citestyle=numeric]{biblatex}
\hypersetup{
    colorlinks = true,
    linkcolor = blue,
    citecolor = teal,
    urlcolor  = magenta
}

\title{Constraint Penalization Method in the Lattice Boltzmann Method (LBM) for Fluid–Structure Interaction}
\author[1]{Tristan Millet}
\author[1]{Erwan Liberge}
\affil[1]{\small LaSIE -  UMR CNRS 7356, Université de La Rochelle}
\date{}
\begin{document}

\maketitle
\begin{abstract}
A constraint penalization method is introduced within the Lattice Boltzmann (LBM) framework to model fluid–structure interactions involving rigid bodies. The proposed approach extends the fictitious domain concept by enforcing the rigid-body motion through a penalization term directly applied to the fluid velocity field, eliminating the need for explicit Lagrange multipliers or interface force computation. This formulation preserves the locality and simplicity of the LBM algorithm while ensuring an implicit coupling between the fluid and solid regions. Numerical experiments demonstrate that the method accurately reproduces rigid-body motion and hydrodynamic interactions with minimal additional computational cost. The method is applied to particle sedimentation, starting with a simple example and progressing to increasingly complex cases.
\end{abstract}
\begin{paragraph}{keywords}
Lattice Boltzmann Method (LBM), Fluid Structure Interaction (FSI), Constraint Penalization, Fictitious Domain, particle sedimentation 
\end{paragraph}

\section{Introduction}

The Lattice Boltzmann Method (LBM) is a mesoscopic numerical approach for simulating fluid flows. Unlike traditional computational fluid dynamics (CFD) techniques—such as finite difference or finite element methods—which discretize the Navier–Stokes equations directly, the LBM models the fluid as a collection of particle populations evolving on a discrete lattice in space, time, and velocity. These populations obey simplified versions of the Boltzmann equation that account for collision and streaming processes. Through a Chapman–Enskog expansion \cite{KRUGER}, the macroscopic Navier–Stokes equations can be recovered, while retaining the advantages of a simple, local, and highly parallelizable algorithmic structure. Due to these properties, the LBM has been successfully applied to particle-laden flows and fluid–structure interaction (FSI) problems~\cite{LADD2001, AIDUN, FENG2004, LIBERGE2024}. 

Several strategies have been developed to model the coupling between fluids and solids in LBM, including the momentum exchange (ME) method~\cite{TAO}, the immersed boundary method (IBM)~\cite{FENG2004}, and volume penalization method~\cite{LIBERGE2024}. All of these approaches share a common feature: the explicit computation of fluid–solid forces, which requires the transfer of information between the fluid solver and the solid. This coupling can significantly slow down simulations and may introduce numerical errors.  

In this work, we focus on adapting the fictitious domain method initially proposed by Glowinski and Patankar \cite{GLOWINSKI1999, PATANKAR2000, GLOWINSKI2001}. The central idea of this approach is to treat the solid region as a fluid and enforce rigid-body motion through Lagrange multipliers. T. Coupez \emph{et al.} \cite{COUPEZ2007, LAURE2005} revisited this concept and compared the enforcement of rigid-body constraints using either Lagrange multipliers or a penalization approach. Their findings indicate that penalization alone is generally sufficient, and that Lagrange multipliers only provide a noticeable increase in accuracy for very fine meshes. To our knowledge, this penalization strategy has so far been implemented exclusively within the finite element framework \cite{JANELA2005}.  

The present study aims to extend this methodology to the LBM framework, offering a potentially efficient and accurate alternative for simulating rigid solids in fluid flows without the explicit computation of interface forces. The first part presents the theoretical context, from the idea at the macroscopic level to its application within the framework of the LBM. Next, five applications are proposed. The first is a simple case of a floating particle in a fluid flow, followed by examples dealing with variations in the shape, number, and density of particles in sedimentation.

\section{Theoretical background}
\subsection{Monolithic Formulation for Fluid--Solid Interaction}

In the monolithic approach to fluid--structure interaction, both the fluid and the solid are treated within a single computational domain, allowing the enforcement of solid constraints without explicitly computing interface forces.

In the fluid domain $\Omega_f$, the incompressible Navier--Stokes equations read
\begin{equation}
\begin{cases}
\rho_f \left( \dfrac{\partial \mathbf{u}_f}{\partial t} + \mathbf{u}_f \cdot \nabla \mathbf{u}_f \right)
= \mathbf{f}_f + \nabla \cdot \boldsymbol{\sigma}_f,\\[1mm]
\nabla \cdot \mathbf{u}_f = 0,
\end{cases}
\end{equation}
with the Cauchy stress tensor decomposed as
\begin{equation}
\boldsymbol{\sigma}_f = -p\,\mathbf{I} + \boldsymbol{\tau}, \qquad 
\boldsymbol{\tau} = 2 \mu_f \mathbf{D}[\mathbf{u}_f], \qquad
\mathbf{D}[\mathbf{u}_f] = \frac{1}{2} (\nabla \mathbf{u}_f + \nabla \mathbf{u}_f^T),
\end{equation}
where $\boldsymbol{\tau}$ is the viscous stress tensor.

In the solid domain $\Omega_s$, rigid-body constraints are enforced either via a Lagrange multiplier $\boldsymbol{\lambda}$ or through penalization of the constraint $\mathbf{D}[\mathbf{u}_s] = 0$:
\begin{equation}
\begin{cases}
\rho_s \left( \dfrac{\partial \mathbf{u}_s}{\partial t} + \mathbf{u}_s \cdot \nabla \mathbf{u}_s \right)
= \mathbf{f}_s + \nabla \cdot \boldsymbol{\sigma}_s,\\[1mm]
\nabla \cdot \mathbf{u}_s = 0,
\end{cases}
\end{equation}
with
\begin{equation}
\boldsymbol{\sigma}_s = -p\,\mathbf{I} + 2 \mu_s \mathbf{D}[\mathbf{u}_s] + \mathbf{D}[\boldsymbol{\lambda}],
\end{equation}
where $\mu_s$ is a penalization viscosity enforcing near-rigid behavior in $\Omega_s$, and $\boldsymbol{\lambda}$ represents the Lagrange multiplier enforcing rigid-body motion.

\medskip

Combining the fluid and solid, the governing equations over the full domain $\Omega = \Omega_f \cup \Omega_s$ can be written as~\cite{COUPEZ2007}\cite{PATANKAR2000}
\begin{equation}
\rho(\mathbf{x}) \left( \frac{\partial \mathbf{u}}{\partial t} + \mathbf{u} \cdot \nabla \mathbf{u} \right)
= \rho(\mathbf{x}) \mathbf{g} + \nabla \cdot \boldsymbol{\sigma}(\mathbf{x}),
\end{equation}
with
\begin{align}
\boldsymbol{\sigma}(\mathbf{x}) &= \boldsymbol{\sigma}_s \, \chi(\mathbf{x}) + \boldsymbol{\sigma}_f \, [1-\chi(\mathbf{x})], \\
\mu(\mathbf{x}) &= \mu_s \, \chi(\mathbf{x}) + \mu_f \, [1-\chi(\mathbf{x})], \\ 
\rho(\mathbf{x}) &= \rho_s \, \chi(\mathbf{x}) + \rho_f \, [1-\chi(\mathbf{x})],
\end{align}
where $\chi$ is the indicator function of the solid domain:
\begin{equation}
\chi(\mathbf{x}) = 
\begin{cases}
1, & \mathbf{x} \in \Omega_s,\\
0, & \mathbf{x} \in \Omega_f.
\end{cases}
\end{equation}

This monolithic formulation allows the treatment of fluid and solid within a single system, avoiding the explicit evaluation of fluid--solid interaction forces. Finally, following the observations of Coupez \cite{COUPEZ2007}, which indicate that Lagrange multipliers only significantly improve accuracy for very fine meshes, we opt for a constraint penalization approach, thus avoiding the additional computational expense of Uzawa iterations loop. 
\subsection{Incompressible Lattice Boltzmann Method: Velocity-Based Approach (D2Q9)}

To solve the incompressible Navier--Stokes equations, we employ a Lattice Boltzmann Equation (LBE) formulation based on the velocity field~\cite{GUO2000,GUO2002}. 
Unlike conventional momentum-based schemes, which evolve the macroscopic momentum $\rho \mathbf{u}$, the velocity-based formulation directly updates the velocity $\mathbf{u}$ and the hydrodynamic pressure $p$. 
This approach allows for a monolithic treatment of regions with different densities, enabling the consistent handling of multi-density flows, which is challenging in conventional LBM formulations.

The evolution of the discrete distribution function $f_\alpha(\mathbf{x},t)$, associated with the discrete velocity $\mathbf{c}_\alpha$, is governed by
\begin{equation}
f_\alpha(\mathbf{x} + \mathbf{c}_\alpha \Delta t, t + \Delta t) - f_\alpha(\mathbf{x}, t)
= -\frac{1}{\tau_f} \left[f_\alpha(\mathbf{x},t) - f_\alpha^{eq}(\mathbf{x},t)\right]
+  F_\alpha \, \Delta t + S_\alpha \, \Delta t, 
\label{eq:lbe}
\end{equation}
where $\tau_f$ is the dimensionless relaxation time, $\Delta t$ is the time step, and $F_\alpha$ represents a forcing term, and $S_\alpha$ a source term. The difference between $F_\alpha$ and $S_\alpha$ will be discussed later. 

\begin{figure}[h!]
\centering
\begin{tikzpicture}[scale=1.2, every node/.style={scale=0.9}]
    \draw[step=1cm,gray!50,thin] (-4.5,-1.5) grid (4.5,1.5);
    \draw[<->] (-4,-1.2) -- (-3,-1.2) node[midway, below] {$\Delta x$};
    \draw[<->] (-4.2,-1) -- (-4.2,0) node[midway, left] {$\Delta x$};
    
    \draw[->, thick] (0,0) -- (1,0) node[at end, right] {$\mathbf{c}_1$};
    \draw[->, thick] (0,0) -- (0,1) node[at end, above] {$\mathbf{c}_2$};
    \draw[->, thick] (0,0) -- (-1,0) node[at end, left] {$\mathbf{c}_3$};
    \draw[->, thick] (0,0) -- (0,-1) node[at end, below] {$\mathbf{c}_4$};
    \draw[->, thick] (0,0) -- (1,1) node[at end, above right] {$\mathbf{c}_5$};
    \draw[->, thick] (0,0) -- (-1,1) node[at end, above left] {$\mathbf{c}_6$};
    \draw[->, thick] (0,0) -- (-1,-1) node[at end, below left] {$\mathbf{c}_7$};
    \draw[->, thick] (0,0) -- (1,-1) node[at end, below right] {$\mathbf{c}_8$};
    
    \filldraw[black] (0,0) circle (2pt) node[below right, xshift=10pt, yshift=-1pt] {$\mathbf{c}_0$};
\end{tikzpicture}
\caption{Discrete velocity set for the two-dimensional nine-velocity (D2Q9) lattice used in the present LBM implementation.}
\label{fig:grille}
\end{figure}
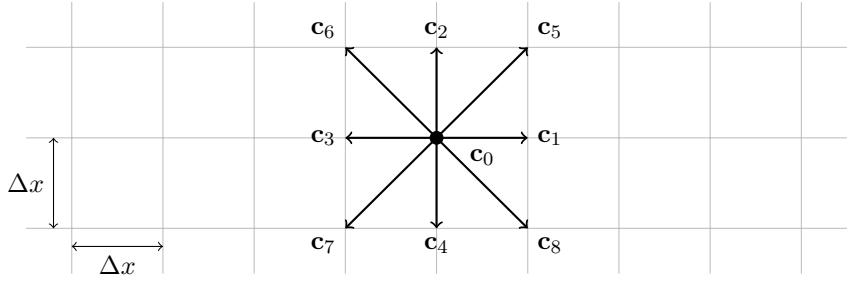

For the two-dimensional nine-velocity (D2Q9) model, the discrete velocity set is illustrated in Fig.~\ref{fig:grille}. 
Each discrete velocity $\mathbf{c}_\alpha$ is associated with a weight coefficient $\omega_\alpha$, defined as
\begin{equation}
\omega_\alpha =
\begin{cases}
4/9, & \alpha = 0,\\
1/9, & \alpha = 1,3,5,7,\\
1/36, & \alpha = 2,4,6,8,
\end{cases}
\label{eq:weights}
\end{equation}
and the lattice speed is given by $c = \Delta x / \Delta t$.

The equilibrium distribution function $f_\alpha^{eq}$ satisfies the following moment constraints:
\begin{equation}
\sum_\alpha f_\alpha^{eq} = 0, \qquad 
\sum_\alpha f_\alpha^{eq}\mathbf{c}_\alpha = \mathbf{u}, \qquad 
\boldsymbol{\Pi}^{eq}=\sum_\alpha f_\alpha^{eq}\mathbf{c}_\alpha\mathbf{c}_\alpha = \mathbf{u}\mathbf{u} + p\mathbf{I},
\label{eq:moments}
\end{equation}
where $p = p_h / \rho$ denotes the normalized hydrodynamic pressure. 

The explicit expression of the equilibrium distribution function reads 
\begin{equation}
f_\alpha^{eq} =
\begin{cases}
-(1 - \omega_0)\dfrac{p}{c_s^2} - \omega_0\dfrac{\mathbf{u}\cdot\mathbf{u}}{2c_s^2}, & \alpha = 0, \\[2mm]
\omega_\alpha \left[\dfrac{p}{c_s^2} + \dfrac{\mathbf{c}_\alpha\cdot\mathbf{u}}{c_s^2} + 
\dfrac{(\mathbf{c}_\alpha\cdot\mathbf{u})^2}{2c_s^4} - 
\dfrac{\mathbf{u}\cdot\mathbf{u}}{2c_s^2}\right], & \alpha \neq 0,
\end{cases}
\label{eq:feq}
\end{equation}
where $c_s = c / \sqrt{3}$ is the lattice sound speed. 

If a force density $\mathbf{f}$ acts on the system (solid or fluid), the corresponding forcing term in the LBE~\eqref{eq:lbe} is given by~\cite{GUO2002FORCE}:
\begin{equation}
\label{eq:force_term}
F_\alpha = \left( 1 - \frac{1}{2\tau} \right) w_\alpha 
\left[ \frac{\mathbf{c}_\alpha - \mathbf{u}}{c_s^2} + \frac{(\mathbf{c}_\alpha \cdot \mathbf{u})}{c_s^4} \, \mathbf{c}_\alpha \right] \cdot \frac{\mathbf{f}}{\rho}.
\end{equation}

The macroscopic velocity and pressure are then recovered as~\cite{ZUANDHE2013,GUO2000}:
\begin{subequations}
\label{eq:macro}
\begin{align}
\mathbf{u} &= \sum_\alpha f_\alpha \mathbf{c}_\alpha + \frac{\Delta t}{2 \rho} \mathbf{f}, \label{eq:macro_u}\\
p &= \frac{1}{1 - \omega_0} \left( \sum_{\alpha \neq 0} f_\alpha c_s^2 - \frac{\omega_0}{2} |\mathbf{u}|^2 \right). \label{eq:macro_p}
\end{align}
\end{subequations}

Through a Chapman--Enskog expansion~\cite{ZUANDHE2013}, the viscous stress tensor is given by
\begin{equation}
\boldsymbol{\tau} = 2 \nu \rho \mathbf{D}[\mathbf{u}] = - \left( 1 - \frac{\Delta t }{2\tau} \right) \boldsymbol{\Pi}^{neq}, 
\end{equation}
where $\nu = c_s^2(\tau - \frac{\Delta t }{2})$ is the kinematic viscosity, and $\boldsymbol{\Pi}^{neq}$ is the non-equilibrium momentum flux tensor defined as 
\[
\boldsymbol{\Pi}^{neq} = \sum_\alpha (f_\alpha - f_\alpha^{eq}).   
\]

To penalize the rigid body constraint , we introduce the following source term in Eq.~\eqref{eq:lbe}:
\begin{equation}
S_\alpha = \frac{\omega_\alpha}{2 c_s^4} \, \mathbf{E} : \left( \mathbf{c}_\alpha \otimes \mathbf{c}_\alpha - c_s^2 \, \mathbf{I} \right),
\end{equation}
where $\mathbf{E} = \alpha \boldsymbol{\Pi}^{neq}$ and $\alpha$ is a real-valued parameter. 
A Chapman--Enskog expansion then shows that the viscous stress tensor is modified as
\begin{equation}
\frac{\boldsymbol{\tau}}{\rho} =  2 \frac{\nu}{1-\alpha \tau} \mathbf{D}[\mathbf{u}]. 
\end{equation}
Hence, the source term introduces an effective viscosity on the solid domain 
\[
\nu_\text{eff} = \frac{\nu}{1 - \alpha \tau}.
\]
For LBM stability, the parameter \(\alpha\) must satisfy
\[
\alpha < \frac{1}{\tau},
\]
so that \(\nu_\text{eff}\) remains positive.
\medskip
\subsection{Collision Strategy}
When multiple solid particles are present in the fluid, short-range repulsive forces are introduced to model \textbf{particle--particle} and \textbf{particle--wall} interactions~\cite{GLOWINSKI2001}\cite{FENG2004}. 

The particle--particle repulsive force exerted on the $i$th particle by the $j$th particle is defined as
\begin{equation}
\mathbf{F}_{\text{col}}^{i,j} =
\begin{cases}
0, & d_{i,j} > R_i + R_j + \delta, \\[1mm]
\frac{C}{\epsilon_p} \, (\mathbf{X}_i - \mathbf{X}_j) \, \left( \dfrac{R_i + R_j + \delta - d_{i,j} }{\delta} \right)^2, & d_{i,j} \le R_i + R_j + \delta,
\end{cases}
\end{equation}
where $d_{i,j} = \|\mathbf{X}_i - \mathbf{X}_j\|$ is the distance between the centers of particles $i$ and $j$, $R_k$ is the radius of particle $k$, $\delta$ is the interaction range, and $\epsilon_p$ is a small positive stiffness parameter, and $C= (\rho_s-\rho_f)g$ a conversion factor.

Similarly, the particle--wall repulsive force is defined as
\begin{equation}
\mathbf{F}_{\text{wall}}^{i,k} =
\begin{cases}
0, & d_{i,k}' > 2 R_i + \delta, \\[1mm]
\frac{C}{\epsilon_w} \, (\mathbf{X}_i - \mathbf{X}_{ik}') \, \left(\dfrac{2 R_i + \delta - d_{i,k}' }{\delta}\right)^2, & d_{i,k}' \le 2 R_i + \delta,
\end{cases}
\end{equation}
where $d_{i,k}' = \|\mathbf{X}_i - \mathbf{X}_{ik}'\|$ is the distance between the particle center and the center of the imaginary particle located on the other side of the $k$th wall, and $\epsilon_w$ is the wall stiffness parameter. The choice of $\epsilon_w$, $\epsilon_p$, and $r$ is discussed in~\cite{GLOWINSKI2001}. 

The total collision force acting on particle $i$ is then obtained by summing over all interacting particles and walls:
\begin{equation}
\mathbf{F}_{\text{col}}^i = \sum_{j \neq i} \mathbf{F}_{\text{col}}^{i,j} + \sum_k \mathbf{F}_{\text{wall}}^{i,k}.
\end{equation}

\subsection{Particle Update}
Given the fluid velocity $\mathbf{u}$ and pressure $p$ at timestep $n$, the particle center-of-mass velocity $\mathbf{U}^n$ and angular velocity $\boldsymbol{\omega}^n$ are computed as
\begin{subequations}
\begin{align}
\mathbf{U}^n &= \frac{1}{M} \int_{\Omega_s^n} \rho_s \, \mathbf{u}^n \, dV, \\
\mathbf{I} \, \boldsymbol{\omega}^n &= \int_{\Omega_s^n} \mathbf{r} \times (\rho_s \mathbf{u}^n) \, dV,
\end{align}
\end{subequations}
where $M$ is the particle mass, $\mathbf{I}$ is the moment of inertia, and $\Omega_s^n$ is the solid domain at time $n$. 

As detailed in Algorithm~\ref{alg:particle_update}, the positions and velocities of particles are updated using a prediction--correction procedure used by Patankar et al.~\cite{PATANKAR2000}. 
\begin{algorithm}[H]
\caption{Explicit update of $i$th particle positions}
\label{alg:particle_update}
\begin{algorithmic}[1]
\STATE Set $\mathbf{X}_i^{n+1,0}= \mathbf{X}_i^n $
\FOR{$k = 1$ to $K$}
    \STATE Prediction 
    \[
    \mathbf{X}_i^{*,n+1,k} = \mathbf{X}_i^{n+1,k-1} + \frac{\Delta t}{K} \frac{\mathbf{U}_i^n + \mathbf{U}_i^{n-1}}{2}
    \]
    \STATE Correction: 
    \[
    \mathbf{X}^{n+1, k }=\mathbf{X}_i^{*,n+1,k}+ \frac{1}{2M}\left(\frac{\Delta t }{K} \right)^2\dfrac{\mathbf{F}_{col}(\mathbf{X}^{n+1, k-1})+\mathbf{F}_{col}(\mathbf{X}^{*,n+1, k})}{2} 
    \]
  
\ENDFOR
\STATE Set $\mathbf{X}_i^{n+1}= \mathbf{X}_i^{n+1,K}$ and update $\Omega_s^{n+1}$. 
\STATE Set
\[
    \mathbf{A}_i^{n+1} =  \frac{2}{(\Delta t ) ^2}\left( \mathbf{X}_i^{n+1}- \mathbf{X}_i^{n} - \Delta t \frac{\mathbf{U}_i^{n}+ \mathbf{U}_i^{n-1}}{2}  \right)
\]
\STATE Set 
\[
F_\alpha^{n+1}=\sum_{i=1}^{N_p} \left( 1 - \frac{1}{2\tau} \right) w_\alpha 
\left[ \frac{\mathbf{c}_\alpha - \mathbf{u}}{c_s^2} + \frac{(\mathbf{c}_\alpha \cdot \mathbf{u})}{c_s^4} \, \mathbf{c}_\alpha \right] \cdot \mathbf{A}_i^{n+1}
\], $N_p$ is the number of particles. 
\STATE Add $F_{\alpha}^{n+1}$ as a forcing term in the LBM (see Eq.~\eqref{eq:force_term}).
\end{algorithmic}
\end{algorithm}
This algorithm ensures that particle positions and velocities are updated consistently with the fluid field, while enforcing short-range repulsive interactions to prevent overlap or wall penetration. 
Moreover, any additional forces acting on the particle (such as gravity) are already incorporated as direct forcing terms in the LBE (see Eq.~\eqref{eq:lbe}). 
 
\section{Numerical Results}
All quantities reported in this study are expressed in lattice units (l.u.), following standard practice in the lattice Boltzmann (LB) literature~\cite{KRUGER}. It means that the lattice spacing and timestep are set to unity, $\Delta x^{*} = \Delta t^{*} = \rho^* = 1$, and therefore the conversion factor for length, time and density and respectively equal to $\Delta x, \Delta t $ and $\rho$. Moreover, in every simulation the $\alpha$ parameter was set to $1/\tau$.  

\subsection{Benchmark: Lateral Migration of a Particle in Plane Couette Flow}
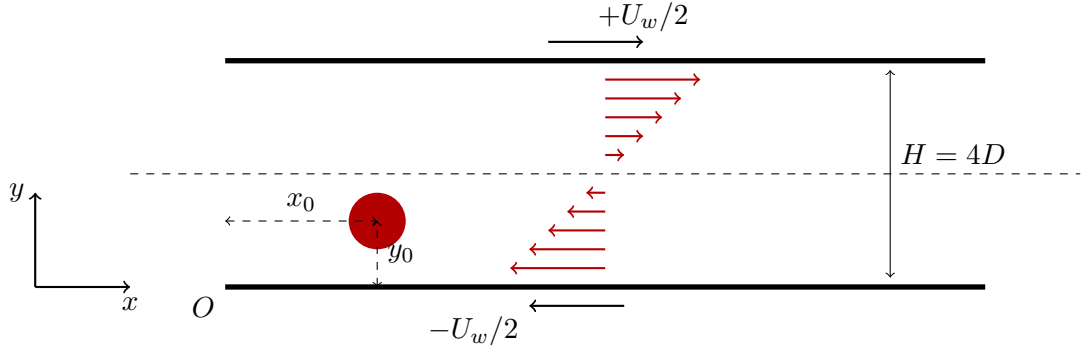
\begin{figure}[h!]
\centering

\begin{tikzpicture}[scale=2.5]

\def\L{4}   
\def\H{1.2} 

\draw[line width=2pt] (0,0) -- (\L,0);
\draw[line width=2pt] (0,\H) -- (\L,\H);

\draw[->, thick] (\L/2-0.3,\H+0.1) -- ++ (0.5,0)  node[above] {$+U_w/2$};
\draw[->,thick] (\L/2+0.1, -0.1) -- ++(-0.5,0) node[below left] {$-U_w/2$};
\def\xo{0.8} 
\def\yo{0.35} 
\def\D{0.3}  
\fill[red!70!black] (\xo,\yo) circle (\D/2);
\draw[dashed]  (-0.5,\H/2) -- (\L+0.5, \H/2);
\draw[<->, dashed] (0,\yo) -- node[above] {$x_0$} (\xo,\yo);
\draw[<->, dashed] (\xo,0) --node[right] {$y_0$} (\xo,\yo);
\node[below left] at (0,0) {$O$};
\draw[->,thick] (-1,0) -- (-0.5,0) node[below] {$x$};
\draw[->,thick] (-1,0) -- (-1,0.5) node[left] {$y$};
\foreach \y in {0.1,0.2,0.3,0.4,0.5} {
    \pgfmathsetmacro{\len}{\y}
    \draw[->,red!70!black,thick] (\L/2,\H/2 + \y) -- ++(\len,0);
}
\foreach \y in {0.5,0.4,0.3,0.2,0.1} {
    \pgfmathsetmacro{\len}{\y}
    \draw[->,red!70!black,thick] ( \L/2, -\y + \H/2) -- ++(-\len,0);
}
\draw[<->] (\L -0.5,0.05) -- node[above right] {$H= 4D$} (\L - 0.5,\H-0.05);
\end{tikzpicture}
\caption{Lateral migration of a neutrally buoyant circular particle in a plane Couette flow. The upper and lower walls move at velocities $\pm U_w/2$, generating a linear shear profile that drives the particle toward the channel mid-plane.}
\label{fig:couette_migration}
\end{figure}

We consider the lateral migration of a neutrally buoyant circular particle (i.e., $\rho_s = \rho_f = 1.0$) suspended in a planar Couette flow. The upper and lower walls are separated by a gap $H$ and move at constant and opposite velocities $\pm U_w/2$, generating a uniform shear rate $\gamma = U_w/H$, as depicted in Figure~\ref{fig:couette_migration}. The computational domain has dimensions $L \times H$, with $H = 4D$ and $L = 100D$, where $D$ denotes the particle diameter. The LBM simulation parameters are : 
\[
(H, \tau, \rho_f, \rho_s, \nu, U_w, T) = (200, 1.0, 1.0, 1.0, 0.01, 1/60, 7 \times 10^5)\]
Initially, the particle is released from position $(x_0, y_0) = (15D, D)$.

\begin{figure}[h!]
    \centering
    \begin{subfigure}[b]{0.55\linewidth}
        \centering
        \includegraphics[width=\linewidth]{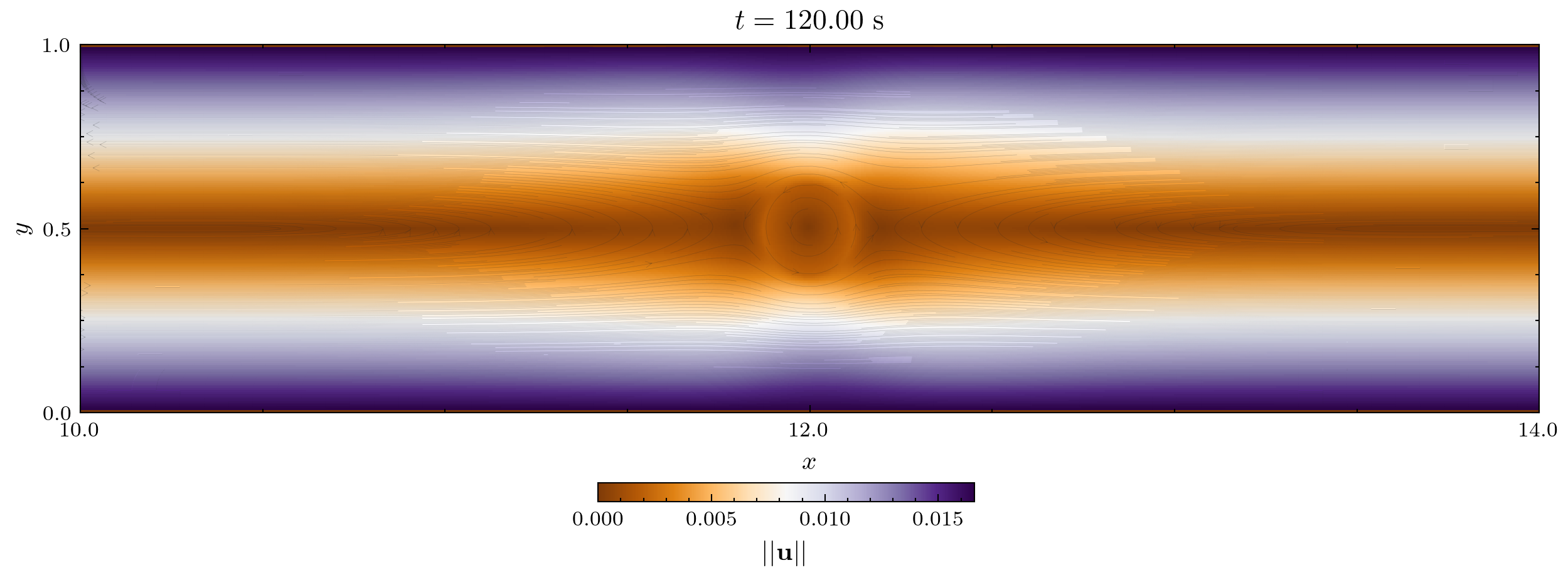}
        \caption{Streamlines and velocity field at $t=120$ s.}
        \label{fig:couette_flow}
    \end{subfigure}
    \hfill
    \begin{subfigure}[b]{0.48\linewidth}
        \centering
        \includegraphics[width=\linewidth]{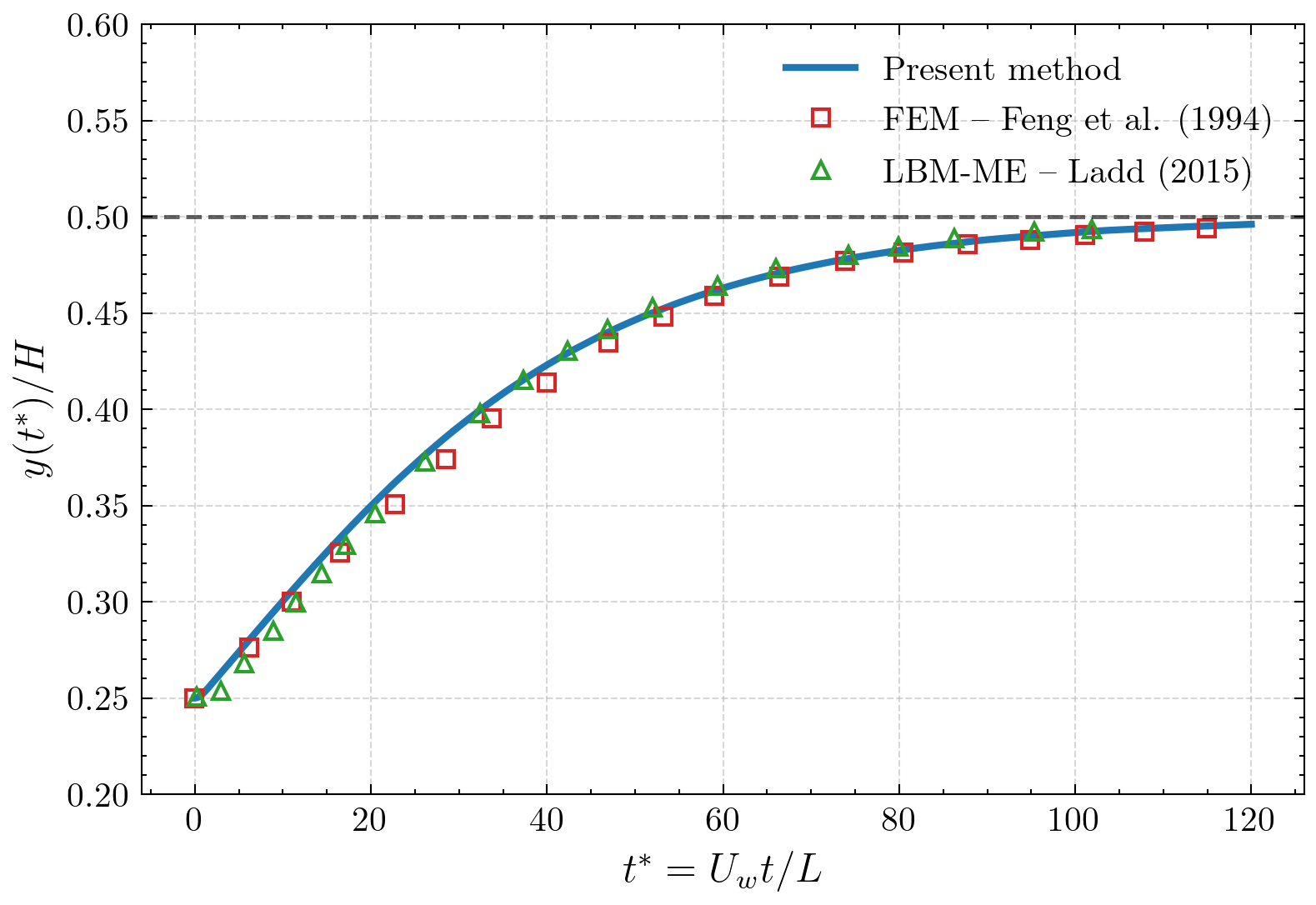}
        \caption{Lateral migration comparison with Feng et al.~\cite{FENG1994SEDI} and Ladd~\cite{TAO}. Dashed line: equilibrium at $y/H = 0.5$.}
        \label{fig:particle_migration}
    \end{subfigure}
    \caption{(a) Couette flow configuration with streamlines, (b) Neutrally buoyant particle migration towards the channel centerline.}
    \label{fig:couette_comparison}
\end{figure}

Figure~\ref{fig:particle_migration} presents the time evolution of the normalized vertical position of the particle center, $y(t^*)/H$, where $t^* = \gamma \, t  $. The results obtained with the present approach are compared with the finite-element simulations of Feng \textit{et al.}~\cite{FENG1994} and the LBM results of Ladd~\cite{LADD1994}. The particle, initially released off-center, gradually migrates toward the channel mid-plane ($y/H = 0.5$), which corresponds to the stable equilibrium position predicted in the literature. A very good quantitative agreement is observed with both reference solutions. Furthermore, Figure~\ref{fig:couette_flow} clearly reveals the rigid body motion within the particle. 

\newpage 
\subsection{Benchmark: Single Particle Settling under Gravity}
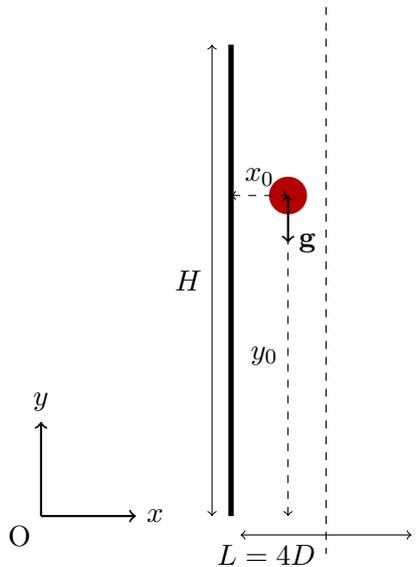
\begin{figure}[h!]
\centering
\begin{tikzpicture}[scale=2.5]

\def\W{1}   
\def\H{2.5} 

\draw[line width=2pt] (0,0) -- (0,\H);
\draw[line width=2pt] (\W,0) -- (\W,\H);
\draw[dashed]  (\W/2, \H+0.2) -- (\W/2,- 0.2);
\def\xo{0.3} 
\def\yo{1.7} 
\def\D{0.2}  

\fill[red!70!black] (\xo,\yo) circle (\D/2);
\draw[->,thick] (\xo,\yo) -- (\xo,\yo-0.25) node[right] {$\mathbf{g}$};

\draw[<->, dashed] (0,\yo) --node[above] {$x_0$} (\xo,\yo) ;
\draw[<->, dashed] (\xo,0) --node[left] {$y_0$} (\xo,\yo );
   
\node[below left] at (-1,0) {O};
\draw[->,thick] (-1,0) -- (-0.5,0) node[right] {$x$};
\draw[->,thick] (-1,0) -- (-1,0.5) node[above] {$y$};
\draw[<->] (-0.1,0) -- node[left] {$H$} (-0.1,\H);
\draw[<->] (0.05,-0.1) -- node[below left] {$L=4D$} (\W-0.05,-0.1);

\end{tikzpicture}
\caption{Sedimentation of a circular particle under gravity in a vertical channel}
\label{fig:disk_initial}
\end{figure}
\begin{figure}[h!]
    \centering
    \begin{subfigure}[b]{0.45\linewidth}
        \centering
        \includegraphics[width=\linewidth]{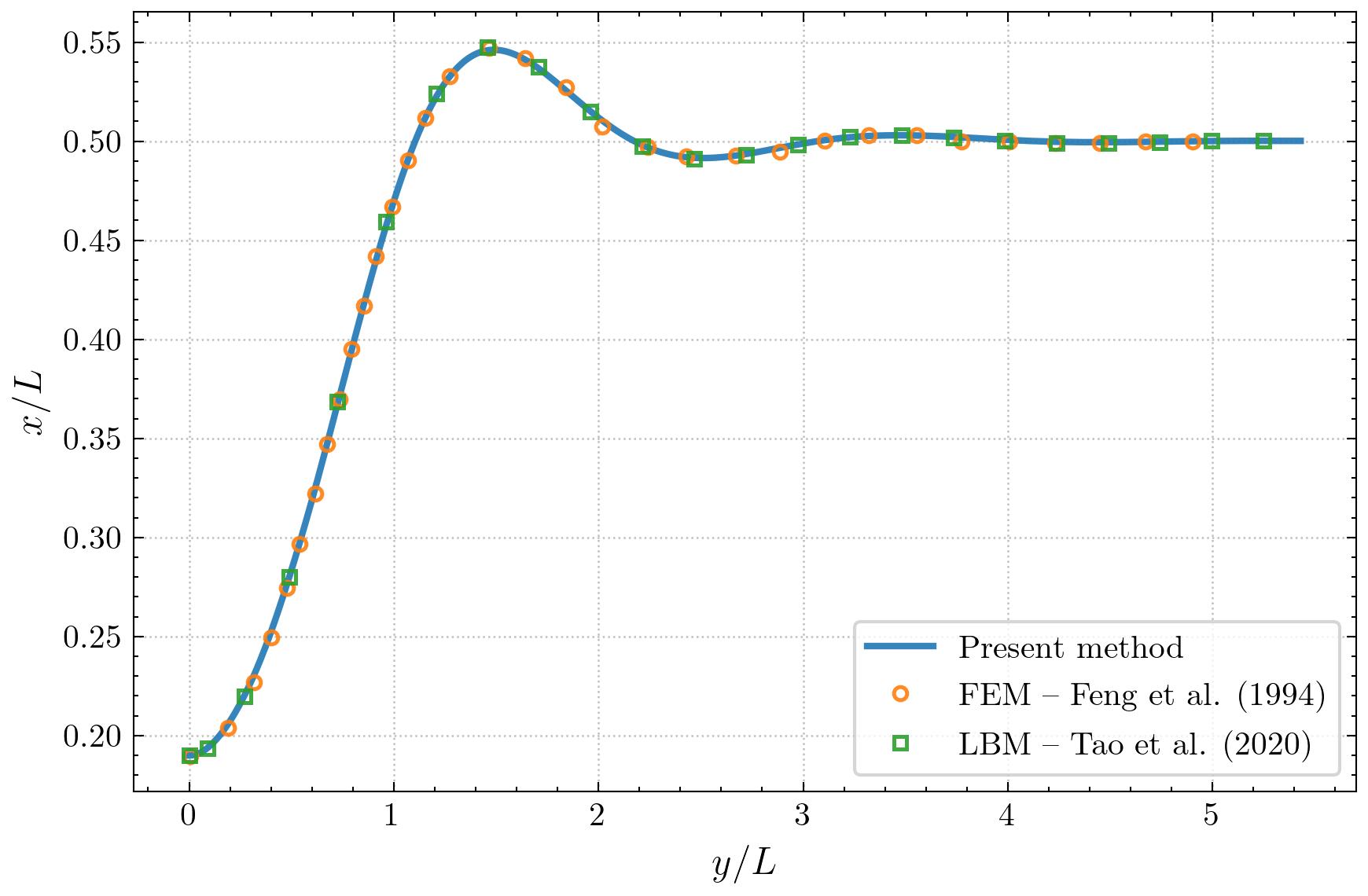}
        \caption{Particle centroid trajectory compared with Tao et al.~\cite{TAO} and Feng et al.~\cite{FENG1994SEDI}}
        \label{fig:particle_settling}
    \end{subfigure}
    \hfill
    \begin{subfigure}[b]{0.45\linewidth}
        \centering
        \includegraphics[width=0.45\linewidth]{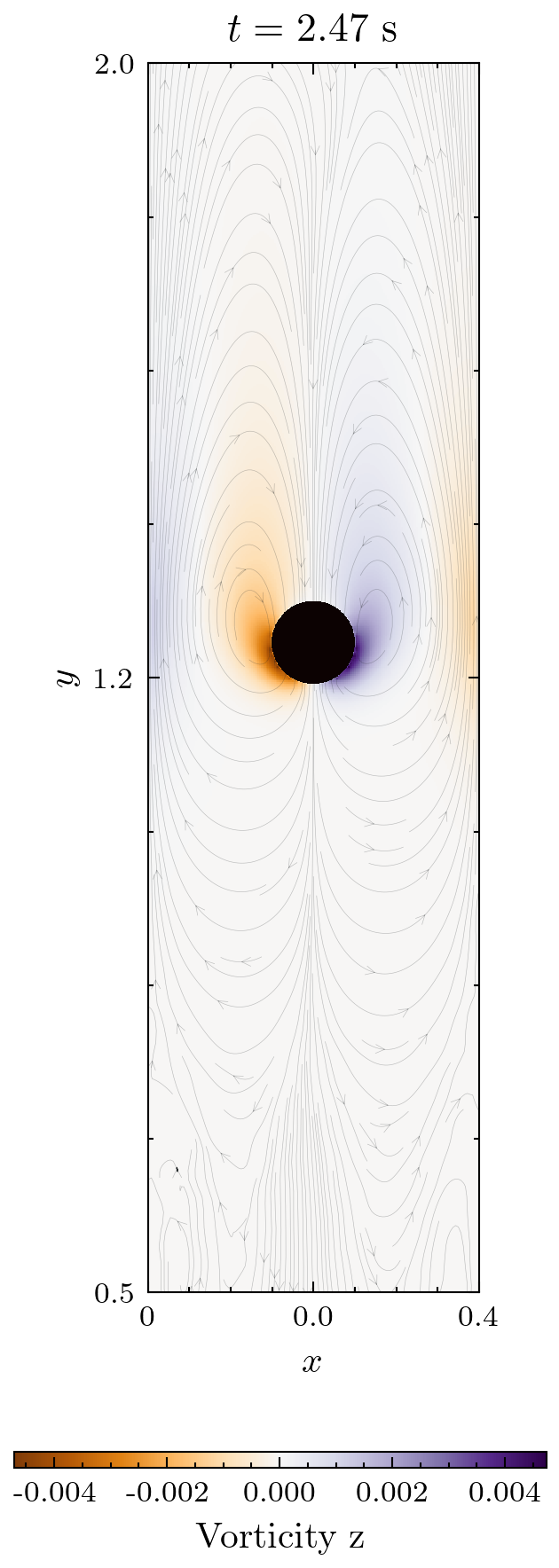}
        \caption{Vorticity field along $z$-axis at $t =  2.47$ s. }
        \label{fig:vorticity_disk}
    \end{subfigure}
    \caption{(a) Particle centroid trajectory, (b) Vorticity field around the settling particle.}
    \label{fig:combined_figures}
\end{figure}

We consider the classical benchmark of a circular particle of diameter $D=0.1$ cm settling under gravity in a vertical channel of size $L \times H= 4D \times 10L $~\cite{TAO} (see Figure~\ref{fig:disk_initial}). The particle has  density $\rho_s = 1.03 \,  \rho_f$ and is initially placed at $(x_0, y_0) = (0.19\, L, 0.8\, H)$ cm.  Gravity acts downward with $g = 980$ cm/s$^2$, and the fluid has kinematic viscosity $\nu = 0.01$ cm$^2$/s. Following \cite{SUN}, the LBM simulation parameters are : 
\[
(L, \tau, \rho_f, \rho_s, \nu, g, T) = (200, 1.0, 1.0, 1.03, 0.001, 1.0, 50000)
\]
which corresponds to $\Delta x = 0.002$ cm, and $\Delta t \simeq 6{,}7 \times 10^{-5}~\mathrm{s}$. The terminal Reynolds number reached by the particle is $Re \approx 8.22$, calculated as $Re = \frac{D \, U_f}{\nu}$ where $U_f$ is the terminal velocity. The vorticity field around the settling particle at $t = 2.47$ s is shown in Figure~\ref{fig:vorticity_disk}.
Particle centroid trajectory and velocity are compared with published benchmarks (see Fig.~\ref{fig:particle_settling})  FEM~\cite{FENG1994SEDI} and BB-LBM~\cite{TAO}. The present results show excellent agreement.
\subsection{Benchmark: Single Elliptical Particle Settling under Gravity}
\begin{figure}[h!]
    \centering

    \begin{subfigure}[b]{0.45\textwidth}
        \centering
        \begin{tikzpicture}[scale=3]
            \def\W{1}   
            \def\H{2.5} 

            \draw[line width=2pt] (0,0) -- (0,\H);
            \draw[line width=2pt] (\W,0) -- (\W,\H);

            \def\xo{0.5}
            \def\yo{2.0}
            \def\ax{0.12}
            \def\ay{0.24}
            \def\thetaa{-45}
            \def\lo{3.5}

            \fill[red!70!black, rotate around={\thetaa:(\xo,\yo)}] (\xo,\yo) ellipse ({\ax} and {\ay});

            \draw[dashed] 
                ({\xo - \lo*\ax*cos(\thetaa)}, {\yo + \lo*\ax*sin(\thetaa)}) -- 
                ({\xo + \lo*\ax*cos(\thetaa)}, {\yo - \lo*\ax*sin(\thetaa)}) ;

            \draw[->] (\xo+0.3, \yo ) arc[start angle=0, end angle=-\thetaa, radius=0.3] node[right] {$\theta$};

            \draw[<->] (\xo,\yo)  -- node[above] {$a$}  
                ({\xo+\ay*cos(\thetaa)}, {\yo-\ay*sin(\thetaa)}); 

            \draw[<->] (\xo,\yo) -- 
                ({\xo-\ax*sin(\thetaa)}, {\yo-\ax*cos(\thetaa)}) node[right] {$b$};

            \draw[->,thick] (\xo-\W,\yo) -- (\xo-\W,\yo-0.25) node[right] {$\mathbf{g}$};

            \draw[<->, dashed] (0,\yo) --node[above] {$x_0$} (\xo,\yo) ;
            \draw[<->, dashed] (\xo,0) --node[left] {$y_0$} (\xo,\yo);
            \draw[dashed] (0,\yo) -- (\W,\yo) ;

            \node[below left] at (-1,0) {O};
            \draw[->,thick] (-1,0) -- (-0.5,0) node[right] {$x$};
            \draw[->,thick] (-1,0) -- (-1,0.5) node[above] {$y$};

            \draw[<->] (-0.1,0) -- node[left] {$H$} (-0.1,\H);
            \draw[<->] (0.05,-0.1) -- node[below left] {$L$} (\W-0.05,-0.1);
        \end{tikzpicture}
        \caption{Schematic representation of the problem.}
        \label{fig:ellipse_initial}
    \end{subfigure}
    \hfill
    \begin{subfigure}[b]{0.45\textwidth}
        \centering
        \includegraphics[width=0.45\linewidth]{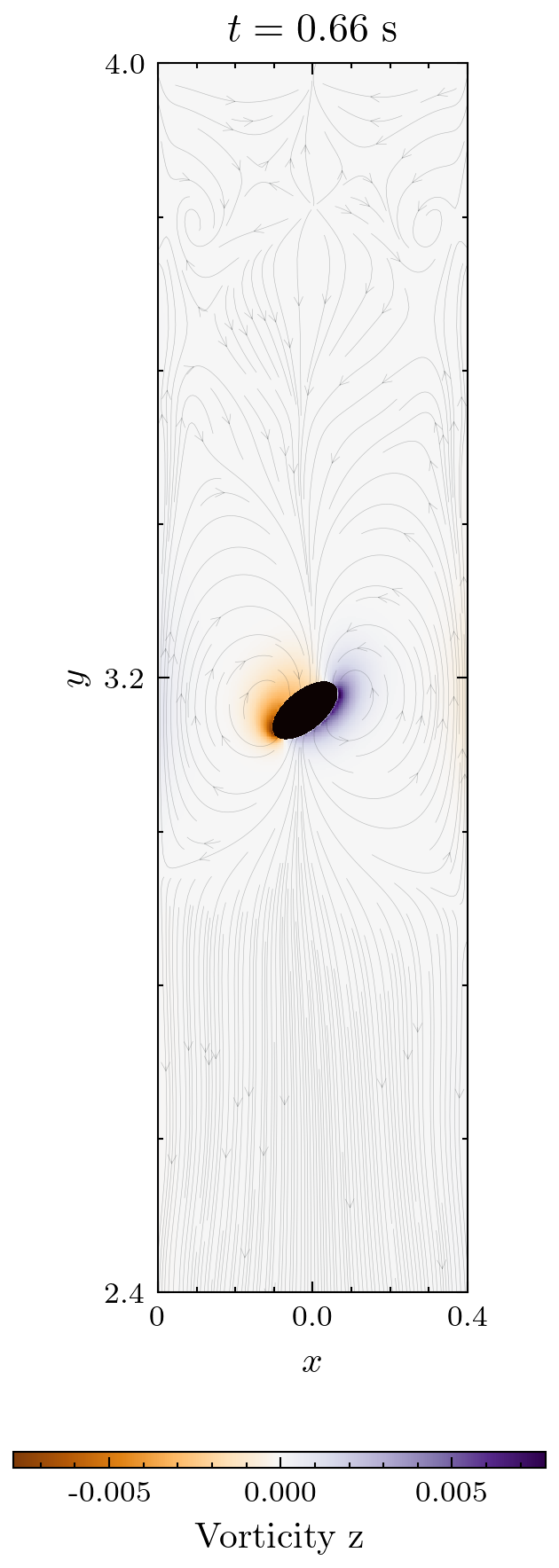}
        \caption{Vorticity field along $z-$axis at $t=\SI{2.6}{\second}$.}
        \label{fig:ellipse_vorticity}
    \end{subfigure}

    \caption{Sedimentation of an ellipse in a vertical channel: (a) schematic geometry and notation; (b) instantaneous vorticity field.}
    \label{fig:ellipse_combined}
\end{figure}
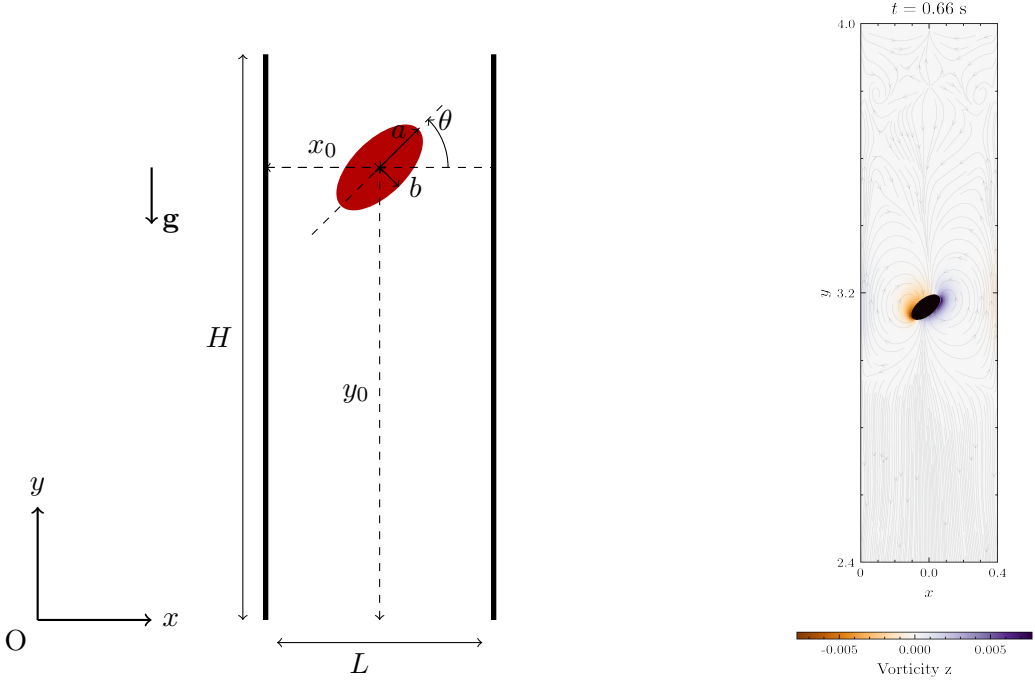

We consider the case of a single elliptical, with major axis $a = 0.05~\mathrm{cm}$ and minor axis $b = 0.025~\mathrm{cm}$, settling under gravity in a vertical channel of size $L \times H =  8a\times 10L$ (see Figure~\ref{fig:ellipse_initial}). 
The particle has a major axis   and density $\rho_s = 1.1 \,\rho_f$. 
It is initially placed at $(x_0, y_0)=(0.5  L, 0.8  H)~\mathrm{cm}$  
Gravity acts downward with $g = 980~\mathrm{cm/s^2}$, and the fluid has kinematic viscosity $\nu = 0.01~\mathrm{cm^2/s}$.
The simulation parameters are:
\[
(L, \tau, \rho_f, \rho_s, \nu, g, T) = (200, 1.0, 1.0, 1.1, 0.01, 30000),
\]
which corresponds to $\Delta x= 0.002~\mathrm{cm}$ and $\Delta t \simeq 0.0006~\mathrm{s}$. The parameters for the wall collision are $\epsilon_w=1$ and $\delta = 1$. The terminal Reynolds number reached by the particle is $Re \approx 11 $, calculated as $Re = \frac{2a \, U_f}{\nu}$ where $U_f$ is the terminal velocity.

\begin{figure}[h!]
    \centering
    \begin{subfigure}[b]{0.48\linewidth}
        \centering
        \includegraphics[width=\linewidth]{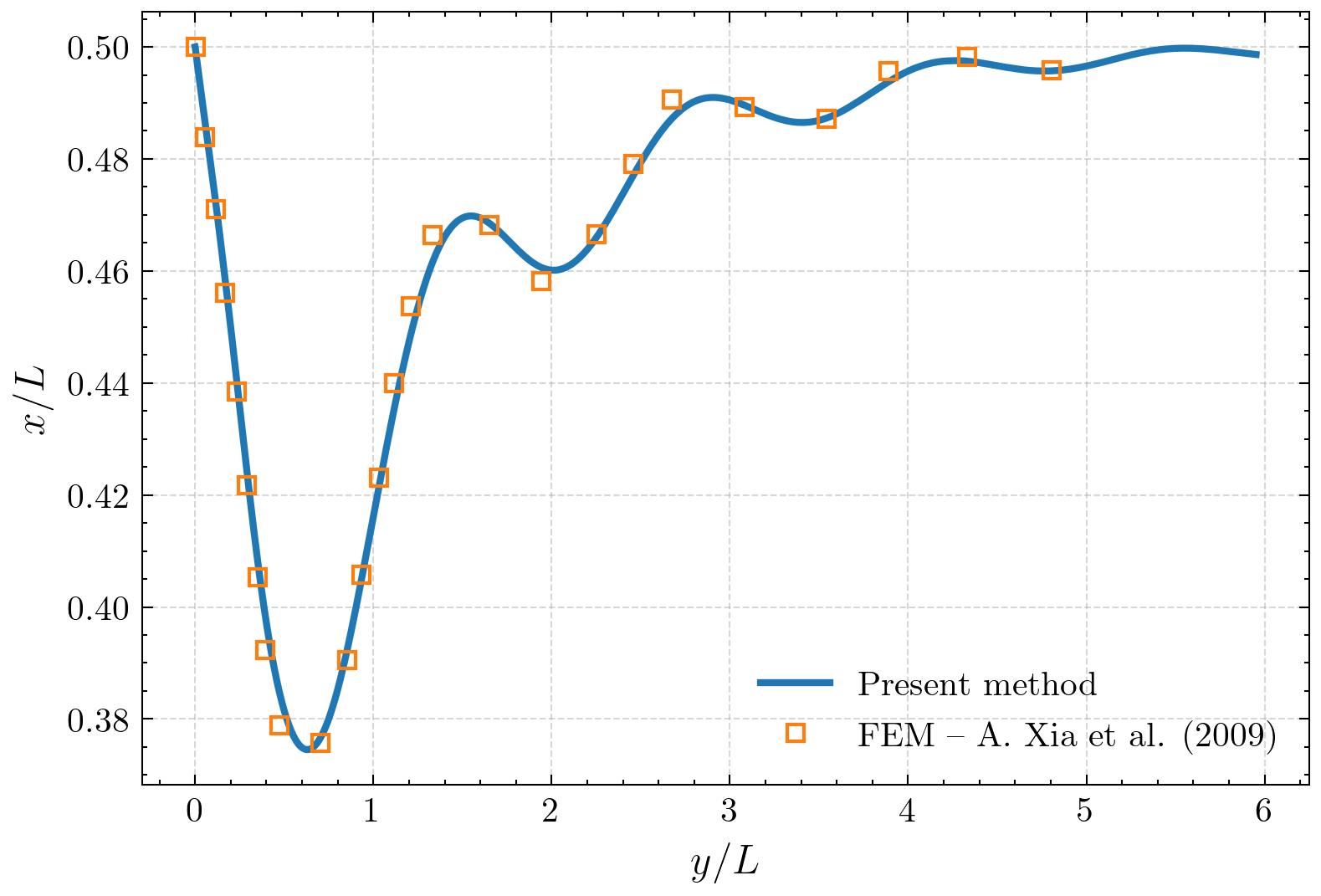}
        \caption{}
        \label{fig:ellipse_x_vs_y}
    \end{subfigure}
    \hfill
    \begin{subfigure}[b]{0.48\linewidth}
        \centering
        \includegraphics[width=\linewidth]{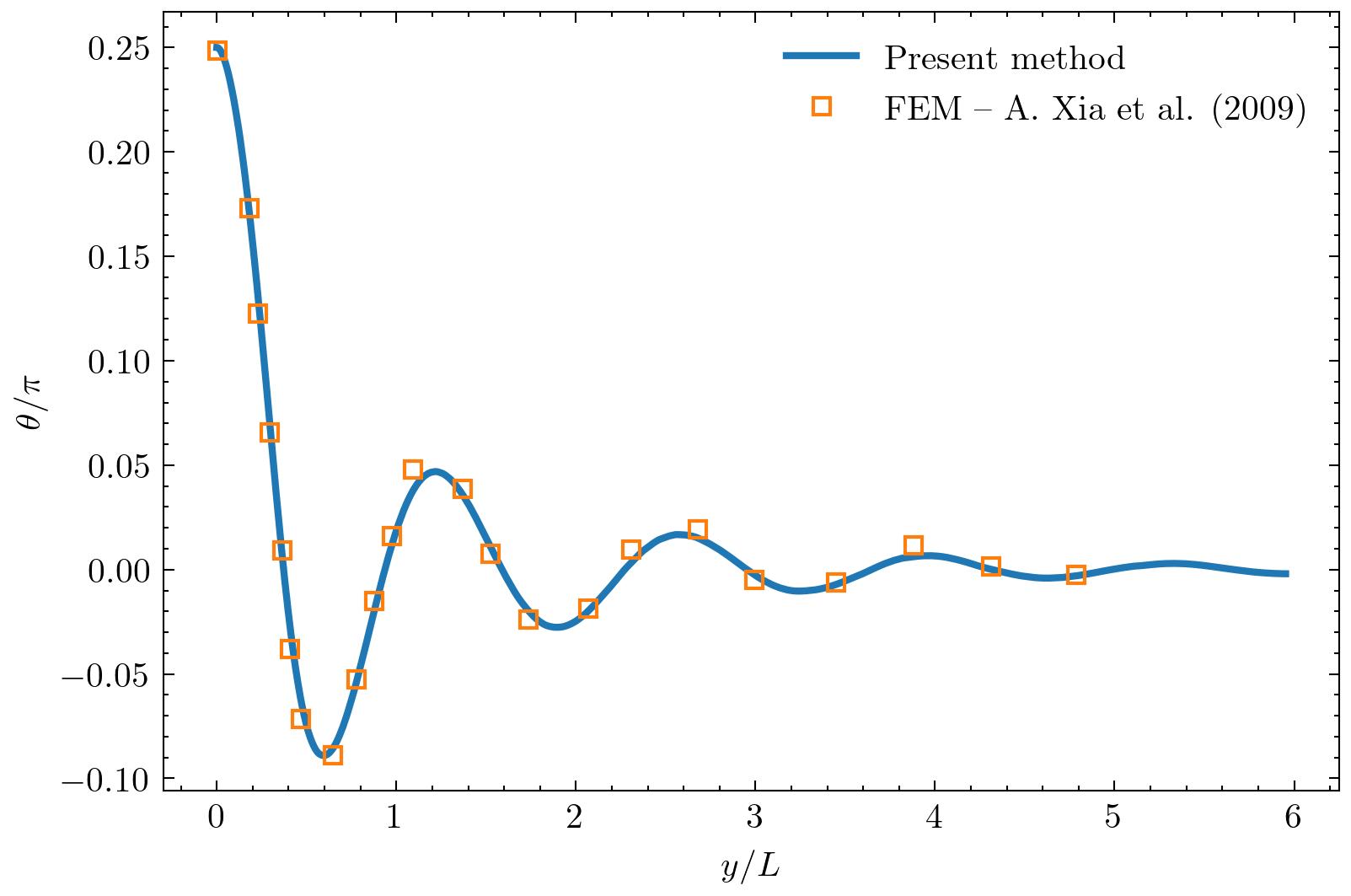}
        \caption{}
        \label{fig:ellipse_theta_vs_y}
    \end{subfigure}
    \caption{Trajectory and orientation of a single elliptical particle settling under gravity. 
    The left panel shows the horizontal position normalized by the domain width, $x/L$, 
    as a function of vertical position $y/L$. The right panel shows the particle orientation 
    $\theta/\pi$ as a function of vertical position $y/L$.}
    \label{fig:ellipse_settling}
\end{figure}

Figure~\ref{fig:ellipse_settling} shows the trajectory of the particle centroid and its orientation angle $\theta$. The particle initially falls under gravity and rotates due to asymmetric hydrodynamic forces and wall interactions. 
The results demonstrate good agreement with published benchmarks for elliptical particles~\cite{XIA_FENG}.

\newpage
\subsection{Drafting--Kissing--Tumbling (DKT) Benchmark}

We consider the drafting--kissing--tumbling (DKT) behavior of two circular particles in a vertical channel. Both particles have the same density, $\rho_s = 1.01\,\rho_f$, and diameter $D= \SI{0.2}{\centi\meter}$. The computational domain is $L \times H = 10D \times 4L$. Gravity acts downward with $g = 980$ cm/s$^2$, and the fluid has kinematic viscosity $\nu = 0.01$ cm$^2$/s. Initially, the particles are positioned at the center of the channel with vertical coordinates $y_0 = \SI{7.2}{\centi\meter}$ (upper particle) and $y_1 = \SI{6.8}{\centi\meter}$ (lower particle). The simulation parameters are :
\[
(L, \tau, \rho_f, \rho_s, \nu, G, T) = (400, 1.0 , 1.0, 1.01, 0.01, 10000)
\]
with $\Delta x = \SI{0.01}{\centi\meter}$ and $\Delta t =1/600\,\si{\second}$. The parameters for the collisions are $\epsilon_p = \epsilon_w = 2$ and $\delta = 1$.

\begin{figure}[h!]
    \centering
   
    \begin{subfigure}[b]{0.49\linewidth}
        \centering
        \includegraphics[width=\linewidth]{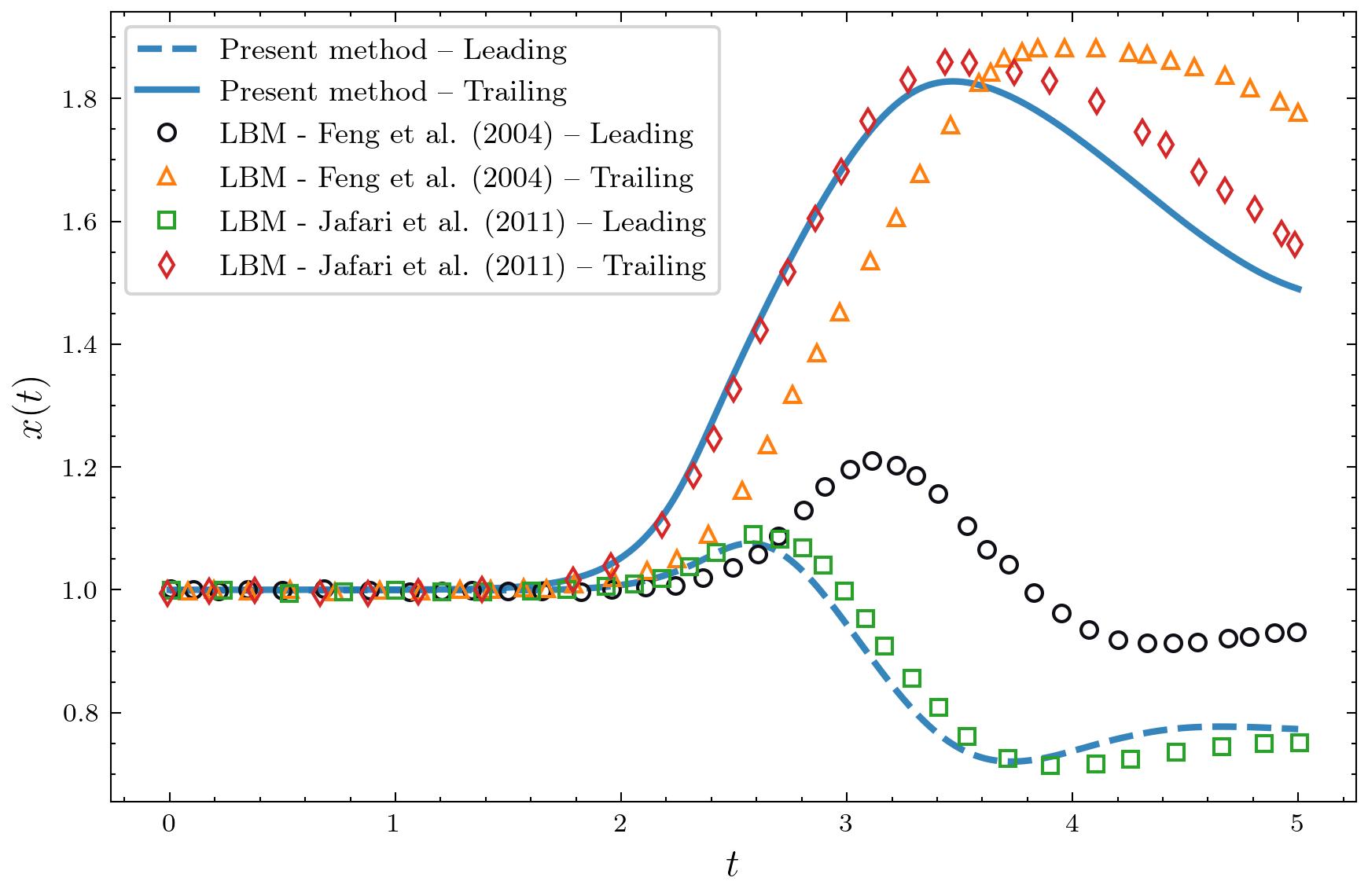}
        \caption{}
    \end{subfigure}
    \hfill
    \begin{subfigure}[b]{0.49\linewidth}
        \centering
        \includegraphics[width=\linewidth]{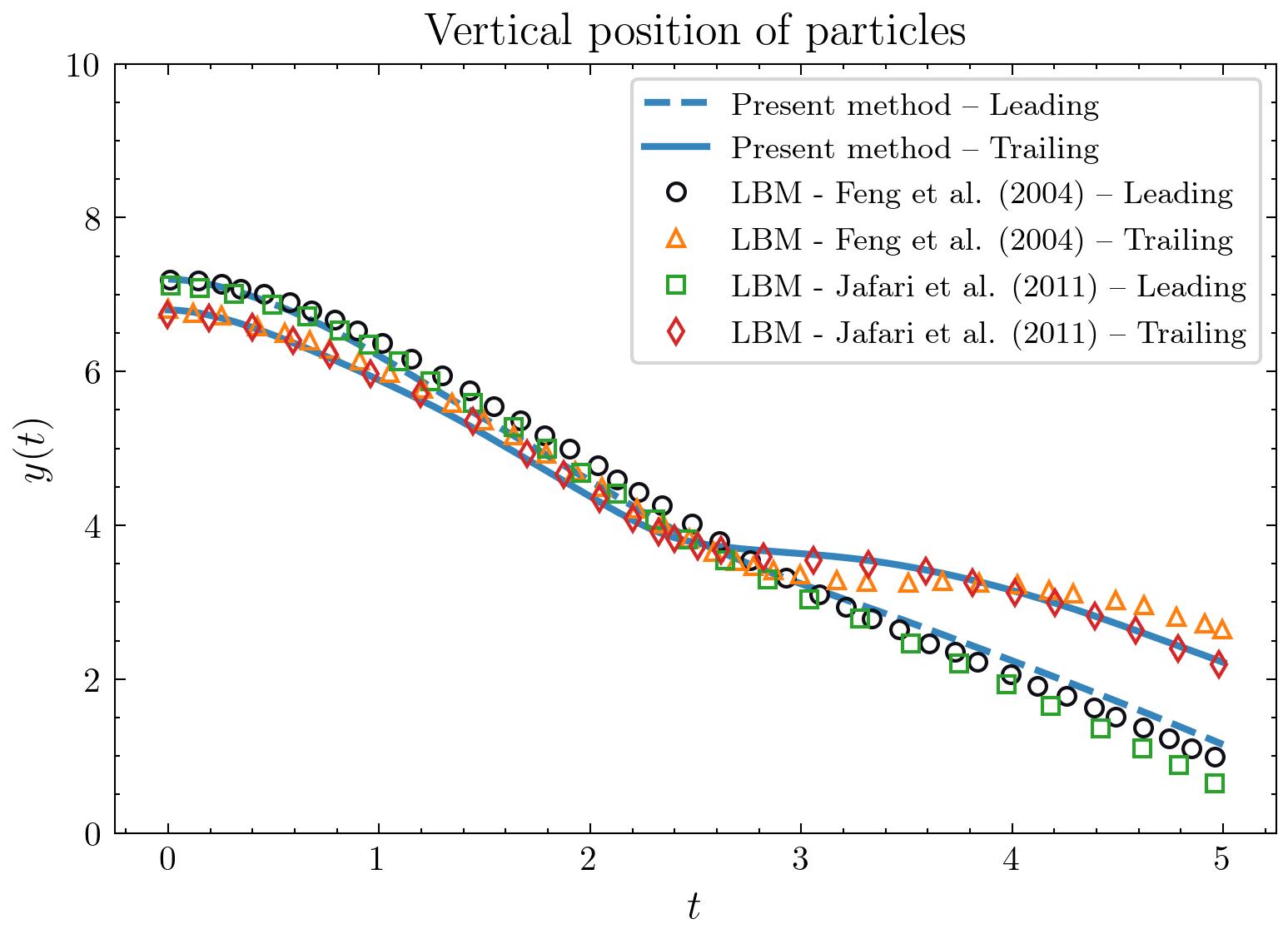}
        \caption{}
    \end{subfigure}
    \caption{Time evolution of $x$ and $y$ coordinates of the particles center of mass. $x$ and $y$ in $\mathrm{cm}$, $t$ in s.}
    \label{fig:dkt}
\end{figure}

\begin{figure}[h!]
    \centering
    \resizebox{0.95\textwidth}{!}{ 
    \includegraphics[width=\linewidth]{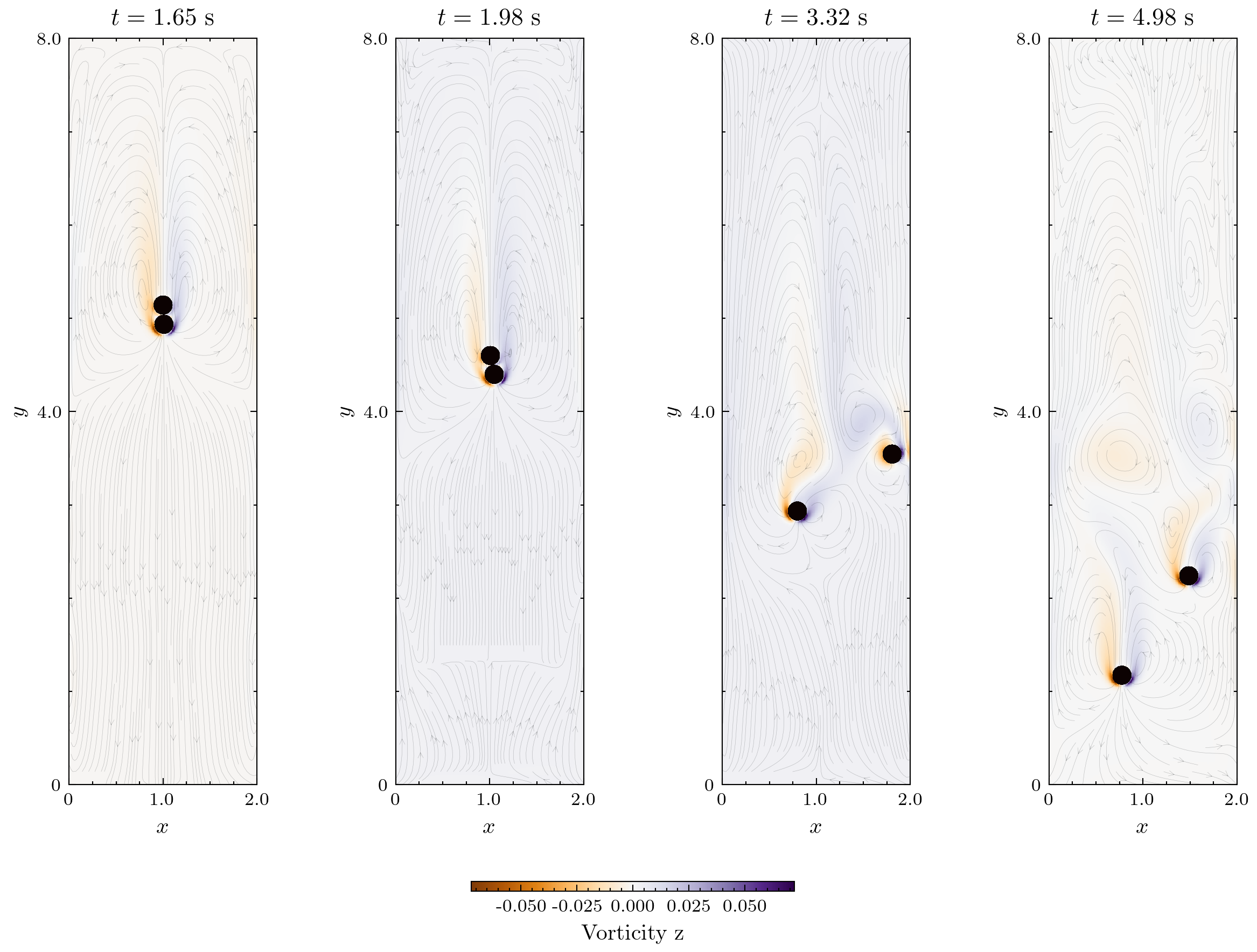}
    } 
    \caption{Vorticity in the $z$-direction and streamlines around the particles at successive times, illustrating the three phases: drafting, kissing, and tumbling.}
    \label{fig:sedimentation_enclosure}
\end{figure}

As expected, the trailing (lower) particle generates a low--pressure wake in which the trailing (upper) particle experiences a reduced drag force and accelerates downward. This initial stage is referred to as \textit{drafting}, clearly observed in Figure~\ref{fig:sedimentation_enclosure} at $t = 1.65\,\mathrm{s}$, where the trailing particle approaches the leading one within its wake. Subsequently, the particles come into close contact (\textit{kissing}) around $t = 1.98\,\mathrm{s}$, forming a temporary elongated configuration along the flow direction. Because this configuration is unstable, by $t = 3.32\,\mathrm{s}$ and $t = 4.98\,\mathrm{s}$, the particles separate and enter in the \textit{tumbling} phase. Figure~\ref{fig:dkt} shows the temporal evolution of the vertical centroid positions of the leading and trailing particles. Our results qualitatively agree with the LBM benchmark simulations of Feng et al.~\cite{FENG2004} and Jafari et al.~\cite{JAFARI2011}.

\vspace{-0.8em} 

\newpage

\subsection{Benchmark: Sedimentation of a Large Number of Particles}
We consider the sedimentation of a large number of circular particles in a closed two-dimensional box, following the setup of Glowinski and Feng ~\cite{GLOWINSKI1999,FENG2004}. The computational domain has a width and height of $\SI{2}{\centi\meter}$, and contains 504 circular particles of diameter $d = \SI{0.0625}{\centi\meter}$. Initially, the particles are arranged in 18 horizontal lines, with 28 particles per line. The horizontal and vertical gaps between particles and between particles and walls are set according to the pattern described in Feng~\cite{FENG2004}, with the first line located at a distance $6d/16$ from the upper wall. The fluid is initially at rest, with density $\rho_f = \SI{1}{\gram\per\centi\meter\cubed}$ and kinematic viscosity $\nu = \SI{1}{\gram\per\milli\second}$. The particle-to-fluid density ratio is $\rho_s = 1.01\rho_f$. The repulsive force between particles has a range $d/16$ and stiffness parameters $\epsilon_p = 2$ and $\epsilon_w =0.5 \epsilon_p$. The simulation parameters are :
\[
(L, \tau, \rho_f, \rho_s, \nu, G, T) = (512, 1.0 , 1.0, 1.01, 0.01, 23600)
\]
which corresponds to $\Delta x \simeq 0.003~\mathrm{cm}$ and $\Delta t \simeq 0.0002~\mathrm{s}$. 
\begin{figure}[h!]
    \centering
    \resizebox{\textwidth}{!}{ 
    \includegraphics[width=\linewidth]{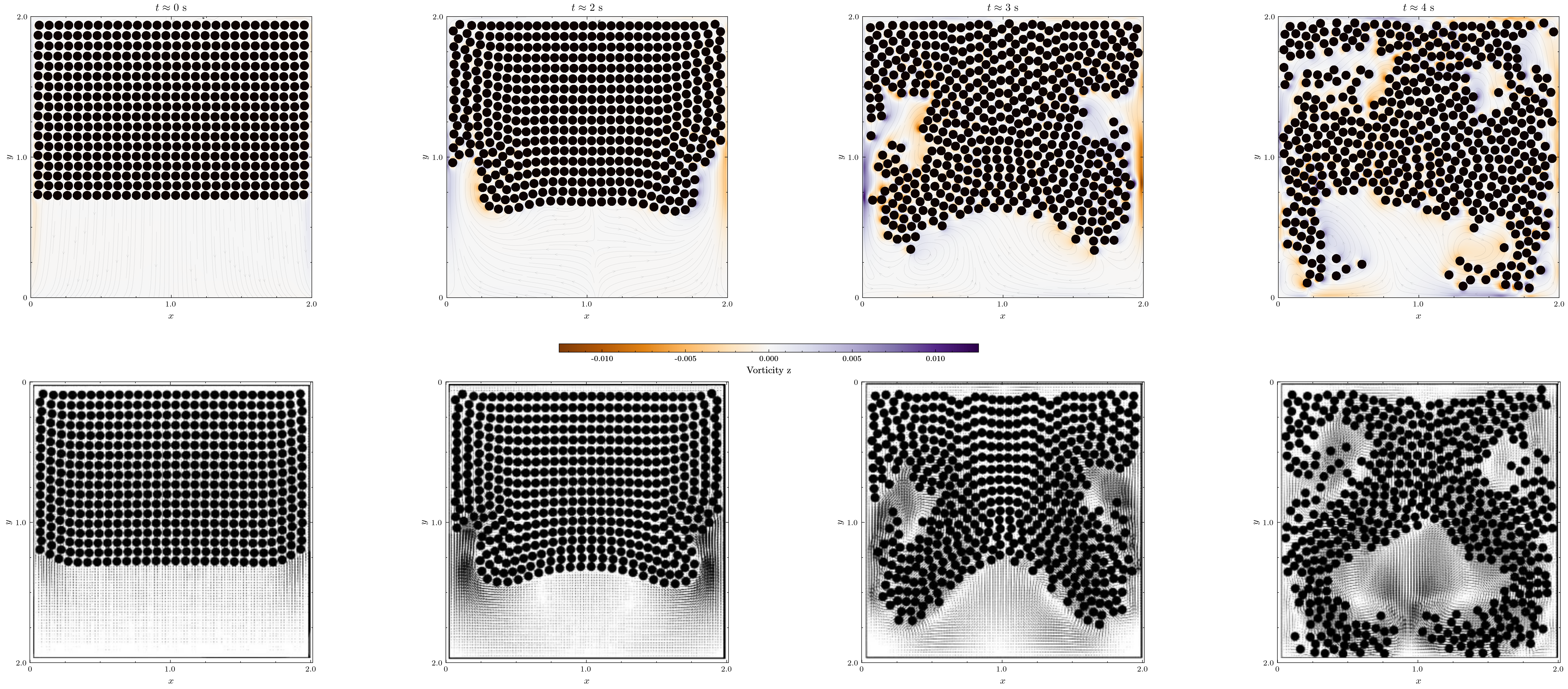}
    } 
    \caption{Comparison of sedimentation of a large number of circular particles in a 2D enclosure. Top row: present simulations; bottom row: results from Feng~\cite{FENG2004}. Each column corresponds to a given time instant, showing particle positions and vorticity along $z-$axis.}
    \label{fig:sedimentation_comparison}
\end{figure}

Initially, all particles start settling uniformly. The walls hinder the closest particles, leading to the creation of side eddies, which split and evolve as the particles are pulled downward (Fig.~\ref{fig:sedimentation_comparison}). This benchmark demonstrates the ability of the present approach to capture collective particle dynamics, hydrodynamic interactions, and Rayleigh–Taylor-like instabilities in dense particulate flows, in good qualitative agreement with Feng~\cite{FENG2004} and Glowinski~\cite{GLOWINSKI1999}.

\section*{Conclusion}
In this work, a constraint penalization method has been developed within the Lattice Boltzmann framework to model fluid–structure interactions involving rigid bodies. The approach extends the fictitious domain concept by enforcing rigid-body motion through a penalization term applied directly to the fluid velocity field. This penalization is added in the LBM framework via a source term such as in the macroscopic equations associated with the LBM scheme, it acts as a divergence of a tensor. This formulation eliminates the need for explicit Lagrange multipliers or force exchange computations, while preserving the locality and efficiency of the LBM algorithm.  The proposed method was shown to accurately capture rigid-body dynamics and fluid–solid coupling without sacrificing computational simplicity, thanks to its monolithic nature. Finally we emphasize that the decision to enforce rigid-body constraints via penalization was motivated by initial attempts using an Uzawa-type algorithm with Lagrange multipliers, which proved difficult to implement successfully. The penalization approach, in contrast, provides a robust and straightforward alternative for enforcing rigid motion within the LBM framework. 

\newpage
\printbibliography[title={References}]

\end{document}